\documentclass[twocolumn,showpacs,preprintnumbers,amsmath,amssymb]{revtex4}

\usepackage{graphicx}
\usepackage{dcolumn}
\usepackage{bm}

\begin{document}


\title{Evaluating Dark Energy Probes using Multi-Dimensional Dark
  Energy Parameters}

\author{Andreas Albrecht}
 \affiliation{Department of Physics, University of
 California at Davis, One Shields Avenue, Davis, CA 9561}
\author{Gary Bernstein}%
\affiliation{Department of Physics \& Astronomy, University of
Pennsylvania, 209 S. 33rd St., Philadelphia, PA 19104
}

\date{August 13, 2006}

\begin{abstract}
We investigate the value of future dark energy experiments by modeling
their ability to constrain the dark energy equation of state.
Similar work was recently reported by the Dark Energy 
Task Force (DETF) using a two dimensional parameterization of the
equation of state evolution.  We examine constraints in a nine dimensional
dark-energy parameterization, and find that the best experiments
constrain significantly more than two dimensions in our 9D space.
Consequently the impact of these experiments is substantially beyond that revealed in the DETF analysis, and the estimated cost per ``impact'' drops
by about a factor of ten as one moves to the very best experiments. 
The DETF conclusions about the relative value of different techniques and
of the importance of combining techniques are unchanged by our
analysis. 

\end{abstract}

\pacs{Valid PACS appear here}
\maketitle

\section{Introduction}

The observed cosmic acceleration requires 
 ``dark energy'' to achieve consistency with the current
cosmological paradigm.  The dark energy must be the dominant component
of the universe today (about 70\% of the energy density), yet an
understanding of its fundamental nature has proved elusive. Many believe
that resolving the mystery of the dark energy will force a
radical change in our understand of fundamental physics. This
expectation has generated great interest in the dark energy 
and widespread enthusiasm for
an aggressive observational program to help resolve this mystery.

Recently, the Dark Energy Task Force (DETF)\cite{Albrecht:2006um}
released a report 
to guide the planning of future dark energy observations. The
DETF used a dark energy ``figure of merit'' (FoM) based on a two-parameter
description of the dark energy evolution in order to produce quantitative
findings. An interesting question is whether using the DETF FoM might lead
to poor choices in shaping an observational program because of its
simplicity.  In particular, could the relative value of two possible
experiments be distorted by the DETF FoM?

We consider this question by
examining an alternative FoM.  We model the dark energy
evolution with a multi-parameter model and formulate a FoM
(the ``D9 FoM'') which gives an experiment credit for {\em any}
constraint it places on the dark energy evolution.
Because theory currently offers little guidance on the functional form
of this evolution, we should seek from experiment any available constraint on its behavior.
We use this alternative FoM to assess many of the same simulated data
sets (or ``data models'') considered by the DETF.  

The DETF dark energy parameters are almost completely unconstrained by
current data which is typically analyzed in smaller parameters spaces
in order to manifest a noticeable impact (see for example
\cite{Spergel:2006hy}). 
In contrast, the DETF approach of evaluating experiments making no
prior assumptions on the cosmic curvature and using a two-parameter model for
the equation of state looks very ambitious ({\em too} ambitious to
some \cite{Linder:2005nh}). Our work shows that the best data models
constrain significantly more than two equation-of-state parameters and
thus their impact was underestimated even using the ambitious
DETF parameterization. As a consequence, we find that the highest
quality large-scale
projects also have a much lower estimated cost per FoM 
improvement than is achieved by the medium scale projects.
 Our work represents only one of several interesting
advances that can reveal impact of dark energy experiments that is
greater than seen by the DETF (see for example
\cite{Zhan:2006gi,Schneider:2006br}). In order to give a more focused
discussion our new FoM is the only significant technical
difference between our methods and those of the DETF.

\section{Dark Energy Parameters}

Following the DETF (and many others) we model the dark energy
as a homogeneous and isotropic fluid. The complete dark energy history
can then be given by the dark energy density today and the ``equation
of state parameter'' (the ratio of the density and pressure of the
fluid) $w(a)$ as a function of time or cosmic scale factor
$a$. 

This paper is about the choice of parametrization of
$w(a)$. The DETF used a standard linear form 
\begin{equation}
  w(a) = w_0 + w_a(1-a)
\label{eqn:w0wa}
\end{equation}
For this work we used a piecewise constant model of $w(a)$
\begin{equation}
  w(a) = -1 + \sum_{i=1}^{N_G} w_i T(a_i,a_{i+1})
\label{eqn:wsteps}
\end{equation}
where the ``tophat function'' $T(a_i,a_{i+1})$ is unity for $a_i>a\geq
a_{i+1}$ and zero otherwise. Any non-zero value
of a $w_i$ gives a deviation from a cosmological constant
($w=-1$).  The DETF linear model is well approximated by a
subspace of our larger space. Here we use $N_G=9$.

The DETF considered the degree to
which $w_0$ and $w_a$ would be constrained by a variety of data models. The
DETF FoM (first used in \cite{Huterer:2000mj})
is given by the reciprocal area in $w_0-w_a$ space
enclosed by the 95\% confidence contour for a given data
model. Correspondingly our 9D FoM is the reciprocal hypervolume
enclosed by the 95\% contour in our larger space.

\section{Methods, $w(a)$ eigenmodes}

Like the DETF we use the Fisher matrix formalism and assume a Gaussian
probability distribution to evaluate the 9D FoM.  
A general (unnormalized) Gaussian probability distribution for
parameters $\vec{x}$  around a central value $\vec{x}_0$ in 
$N$ dimensions is given by 
\begin{equation}
\exp \left\{ {\left( { - \Delta \vec x{\bf{F}}\Delta \vec x} \right)/2} \right\}
\end{equation}
were $\Delta \vec{x} = \vec{x} - \vec{x}_0 $. 
The $N$ eigenvectors $\vec{f_i}$ of ${\rm \bf F}$ give the directions of the
principle axes (or ``principle components'') of the error ellipsoid,
and the width of the error ellipsoid 
along axis $\vec{f_i}$ is given by $\sigma_i$, the inverse square root of the
$i$-th eigenvalue of ${\bf F}$.   The $\vec{f_i}$ describe the
independently measured ``modes'' of $w(a)$ in the 9D space.

We only calculate ratios of FoM's so one can define the FoM
as $\prod_i \sigma_i^{-1}$ (additional constant factors will drop out). 
Depending on whether ${\bf F}$ is defined in the DETF or 9D space of
$w$ parameters, this formula gives 
the DETF or the 9D FoM.  We use many DETF data
models, but we exclude models of galaxy-cluster data, since the
extension of the DETF calculations to our scheme is not
straightforward, especially as regards systematic error estimates.

\begin{figure}
\includegraphics[width=0.5\textwidth]{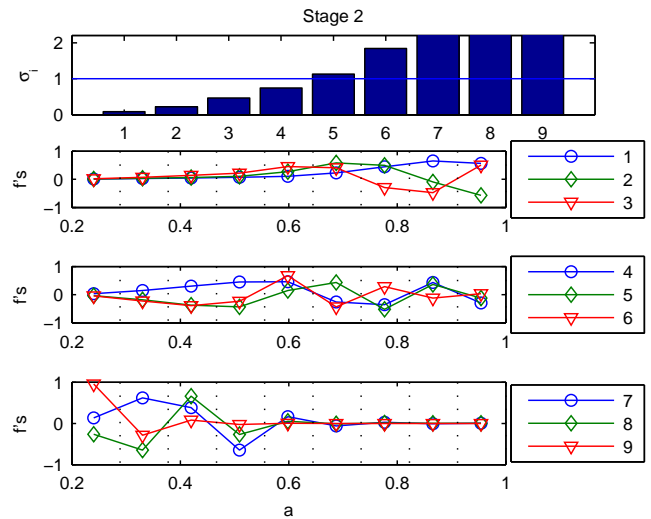}
\caption{\label{fig:Eigs2} Projected impact of experiments currently
 underway (DETF Stage 2) in the 9D space of $w$ parameters: 
 The upper panel gives the errors
  $\sigma_i$ in increasing order. The lower panels
  give the corresponding independently measured modes of $w(a)$ (the
 $\vec{f}_i$).  
The nine markers on each curve give the nine components of each
  $\vec{f}_i$ positioned at $\bar{a}_j \equiv \sqrt{a_ja_{j+1}}$. The
  dotted vertical lines show the values of $a_j$.}  
\end{figure}

\begin{figure}
\includegraphics[width=0.5\textwidth]{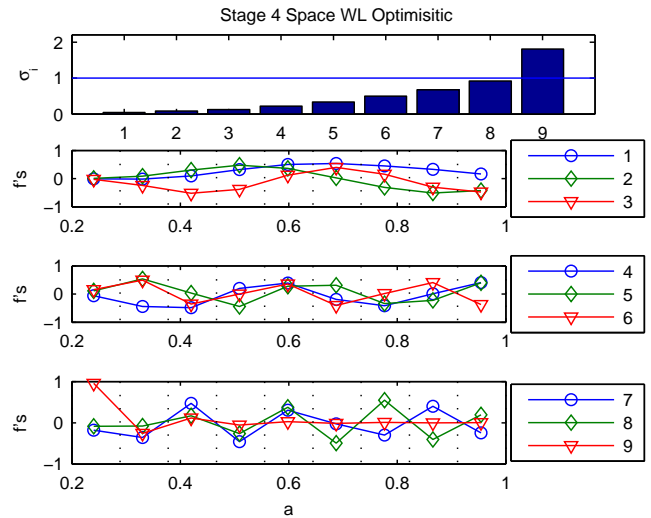}
\caption{\label{fig:Eigs4} The impact of the DETF 
``Stage 4 Space Weak Lensing Optimistic'' data model:  This is a much
  higher quality data set vs Stage 2.  More modes are well 
measured and the well measured modes reach to higher redshift (lower
$a$). This plot has the same format as Fig. \ref{fig:Eigs2}
}
\end{figure}

Each data model corresponds to its own Fisher matrix ${\rm \bf F}$ from
which the the $\vec{f}_i$'s, $\sigma_i$'s and the FoM can be
calculated. Figure \ref{fig:Eigs2} gives the errors 
$\sigma_i$ and modes $\vec{f}_i$ corresponding to the DETF ``Stage
2'' data minus clusters. Stage 2 is  the DETF projection of the data
upon completion of existing projects.
The DETF defines major longer term projects as ``Stage 4'' and
smaller faster future programs as ``Stage 3''. Figure \ref{fig:Eigs4} gives  
the same information for a particular Stage 4 data model. None of
the DETF data models are powerful enough to constrain all nine
directions in our parameter space. This is reflected in large
values of $\sigma_i$ for larger $i$.  We do not trust our formalism to
give meaningful answers in directions that place  
bounds weaker than unity on the $w_i$'s, and such weak constraints are not
likely to enlighten us about dark energy. We handle this issue by
re-setting all $\sigma_i >1$ to unity when we calculate the FoM. That
way changes in $\sigma_i$ in directions where the method is not
trusted do not contribute to changes in the FoM.   

The 9D FoM's reported here use grids $\{a_i\}$ with minimum redshift
$z_{min} \equiv 1/(a_{min}-1)=0.01$,  $z_{max} = 4$ with 
the grid uniform in $a$. We
considered a variety of different grids (varying the spacing pattern,
redshift range and total grid points $N_{G}$)\cite{EPAPS}.
The final grid was chosen to give the highest FoM's which means we
use the parameters which are best measured by the model data
sets. Increasing $N_{G}$ beyond the point where the well measured
modes are well resolved does not significantly change the FoM's or the
shapes and $\sigma_i$'s of the well measured modes.  We find this
convergence under increasing $N_{G}$
an attractive feature of our
parameterization and we found $N_{G}=9$ large enough to achieve
convergence.

Independently measured modes of $w(a)$ (the $\vec{f}_i$'s) reveal
interesting details about each data model.  The fact that the best 
measured Stage 2 $\vec{f}_i$'s (shown in Fig. \ref{fig:Eigs2}) approach zero for $a<0.5$ (larger redshifts)
but the corresponding Stage 4 modes (in Fig. \ref{fig:Eigs4}) do not 
illustrates the deeper redshift reach of the Stage 4 data.
Other specific differences among data models can be understood
by inspecting the plots in \cite{EPAPS}. 

Our form for $w(a)$ has been
used previously by others applied to different data
models\cite{Huterer:2000mj,Huterer:2002hy,Knox:2004vw,Linder:2005ne}
(see also \cite{Riess:2006fw}).
To the extent 
that comparison is possible our work is consistent with these earlier
papers.  In particular, the claims in \cite{Linder:2005ne} that there
are only two measurable $w$ parameters stems from a different formal 
definition of what it means to measure a parameter usefully.  
The choice we make here is best suited to our purpose, which is to make a
direct comparison with the DETF methods. 

There are some small technical differences
between our calculations and those in the DETF report.
When we use two supernova data models in combination they share the
same nuisance parameters (except for the photometric redshifts). 
The DETF calculations keep the nuisance parameters separate.  The
PLANCK prior is the same one used by the DETF, but the Fisher matrix is
expressed in the variables 
$\{n_s, \delta_\zeta, \omega_m, \omega_b,\theta_s\}$  
where we use 
$ln(\theta_s) = -0.252ln(\omega_m) - 0.83ln(\omega_B) -
ln(D_A^{co}(a*)) + 8.2094 $ (from \cite{Hu:2004kn})
These parameters are defined in the DETF report.
Some other modest differences stem from the shortcomings of the
transfer function formalism discussed in section 9.2 of the DETF 
appendix. These small differences from the DETF
calculations are included in all the FoM's presented  here (DETF and 9D) so
our comparisons of FoM's will reflect the different $w$
parameter choices only. 

\section{Results and Interpretation}

\begin{figure*}
\includegraphics[width=0.9\textwidth]{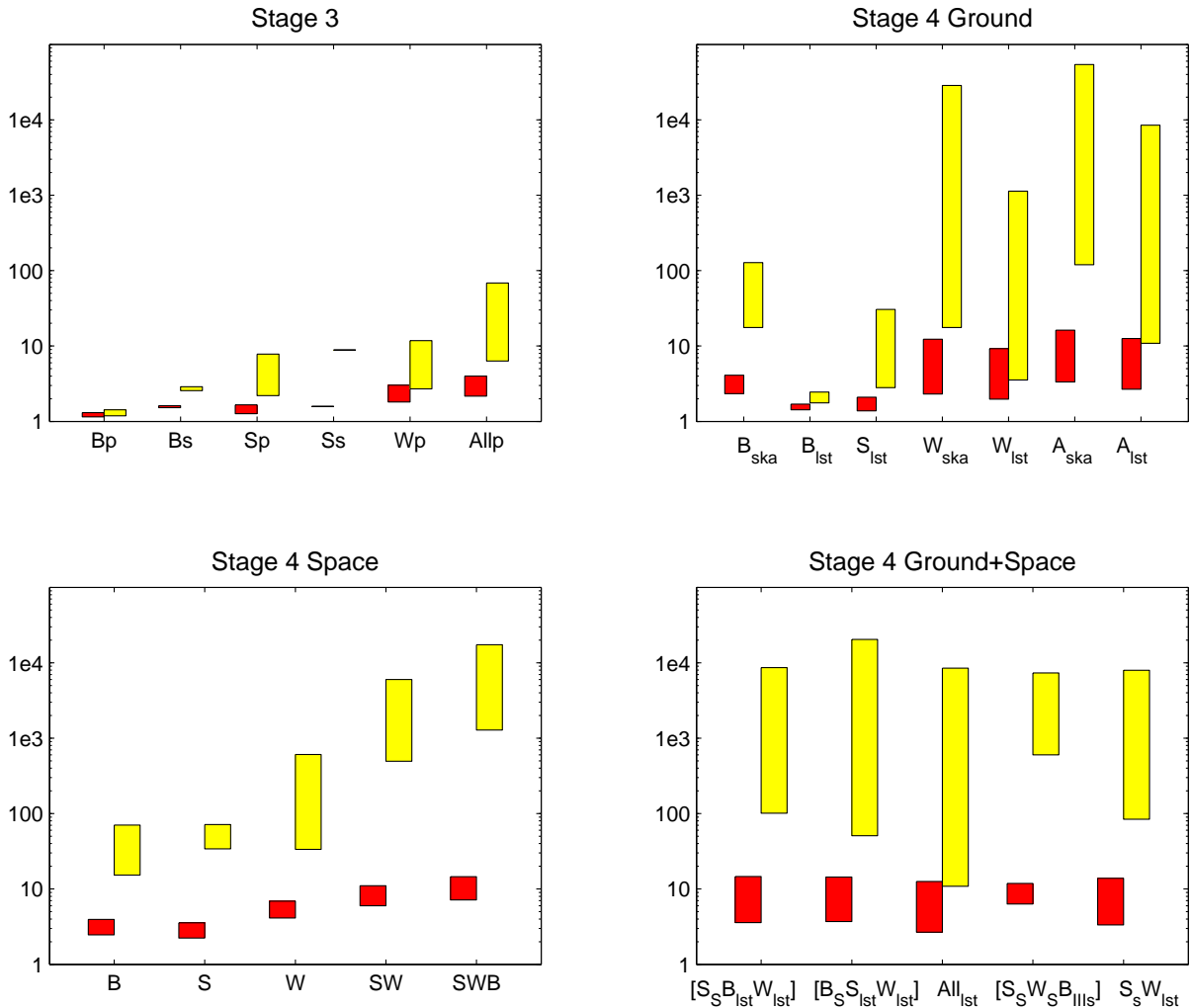}
\caption{\label{fig:DETFbars} Figure of merit improvements vs Stage 2
  for DETF data models: 
  Dark bars show the DETF FoM. To the right of each dark bar is a
  light bar that reflects the 9D FoM for the identical data
  model. The 9D FoM registers a much greater impact from each data
  model, and also shows a much greater improvement at Stage 4 vs Stage
  3. Each panel corresponds to one of the four main
  bar charts in the DETF 
  report (B = ``Baryon Oscillations'', S = ``Supernovae'', W = ``Weak
  Lensing'', A = ``All'' see pp. 16-20 of \cite{Albrecht:2006um}).The x-axis
  labels are similar to those on the DETF plots, but 
  abbreviated. } 
\end{figure*}

The DETF present their main results in four bar charts showing the
FoM ranges for particular data models. The values are given as ratios
to the Stage 2 FoM so that progress beyond Stage 2 can be read
directly, with increasing progress corresponding to larger values
along the y-axis. The four panels in Fig. \ref{fig:DETFbars} show
equivalent plots (using the same DETF data models minus clusters). The
dark bars show the DETF FoM and the light bars show the 9D FoM.  The
9D FoM shows much larger values for all the strong data models. 

One can see that good
combinations of Stage 3 data give 9D FoM improvements of an order of magnitude
and strong Stage 4 data combinations give 9D FoM
improvements of three orders of magnitude or more.  Compared with
the DETF results (about half an order of magnitude to Stage 3 and one
order of magnitude to Stage 4) this is not only a strong showing
overall, but specifically the 9D FoM exhibits a greater improvement
factor going from Stage 3 to Stage 4 as compared with the DETF FoM.
Given the ballpark costs quoted (but not independently verified) by
the DETF of a few $\times \$10^{7} $ for Stage 3 and 
0.3-1 $\times \$10^9 $ for Stage 4, our work indicates that good Stage 4
projects tend to be much more cost effective (about ten times better in \$ per FoM increment)
than Stage 3 projects.  The opposite seems to be the case when using
the DETF FoM, but our work shows that this is only because the 2D DETF
parametrization prevents the better experiments from showing their
full capabilities.

The DETF FoM is constructed in a 2-D parameter
space.  To build intuition, consider the following effective ``reduction to 2-D'' of the 9D FoM:
\begin{equation}
 {\cal F}_{2}  \equiv  {\cal F}_{9}^{2 / D_{e}}  
\label{eqn:Dreduce}
\end{equation}
Here ${\cal F}_2$ and ${\cal F}_9$ are the reduced and regular 9D FoM's
respectively.  One can think of $ {\cal
  F}_{2} $ as 
the product of only two $1/\sigma_{\rm eff}$'s where $\sigma_{\rm eff}$ is a
suitably defined geometric mean of the $\sigma_i$'s.  The ``effective
dimension'' $D_{e}$ can be thought of as the number of directions
constrained by the data model in 9D parameter space.  

\begin{figure*}
\includegraphics[width=0.9\textwidth]{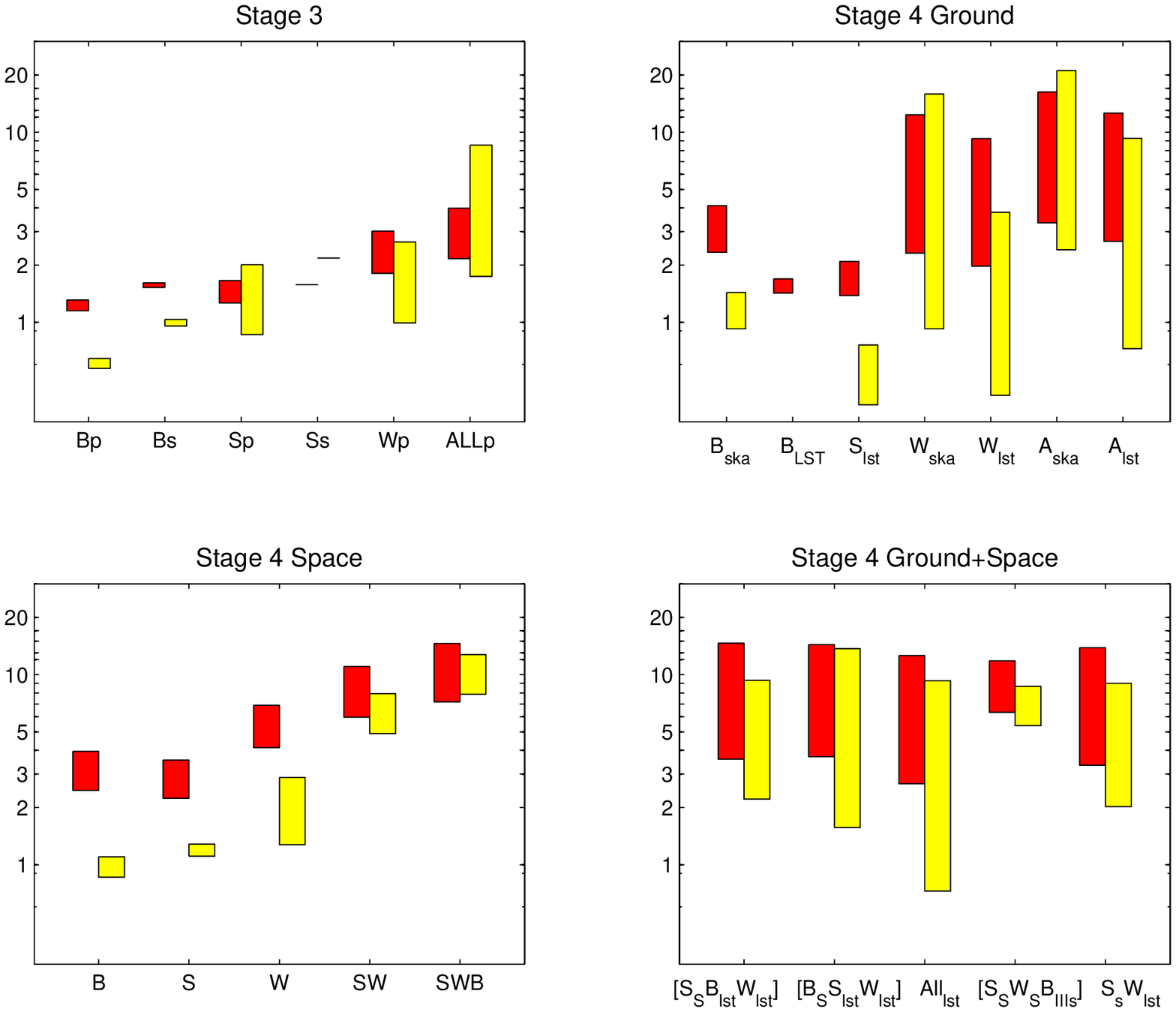}
\caption{\label{fig:DETFbarsReduced} Rescaled FoM's: This is identical
  to  Fig. \ref{fig:DETFbars} except all the 9D FoM's have been
  scaled according to Eqn. \ref{eqn:Dreduce}.  We use $D_e = 2.5$ for
  Stage 2, $D_e = 3$ for Stage 3, $D_e = 4$ for Stage 4 pessimistic
  and $D_e = 4.5$  for Stage 4 optimistic.  These
  plots suggest that the scaling gives a reasonable account of the
  impact of measuring more parameters in 9D vs 2D spaces demonstrated in
  Fig. \ref{fig:DETFbars}. }
\end{figure*}

Figure \ref{fig:DETFbarsReduced} is the same as Fig
\ref{fig:DETFbars} except Eqn. \ref{eqn:Dreduce} has been applied to
all the 9D FoM's. Values of $D_e$ were assigned to the data models
(details in the caption) to create an approximate match (by eye)
between rescaled 9D and DETF FoM's.  Inspection of
Fig. \ref{fig:DETFbarsReduced} suggests that the rescaling accounts
for may of the differences between the FoM's.  A  refined
assignment of $D_e$'s gradually increasing with improving data
models would give an even better account of the differences (but 
still would not match up every detail).  Successful matching of the DETF
and 9D FoM's after rescaling implies that $D_e$ directions in the 9D
space are measured as well as the two DETF parameters.  This applies
only in an average 
sense: The best 9D modes are measured much better than the best DETF
parameter, others are less well measured. 
We present this rescaling to
give intuition about the similarities and differences between the DETF and 9D
FoM's.  The similarity of the DETF and 9D bars in
Fig. \ref{fig:DETFbarsReduced} help one see how the detailed DETF 
conclusions about the relative value of different techniques and
combinations of techniques are unchanged by our analysis. However, we
do not expect Eqn. \ref{eqn:Dreduce} to offer a universal
relationship between the two FoM's that extrapolates to all
possible data models.  

\begin{figure}
\includegraphics[width=0.45\textwidth]{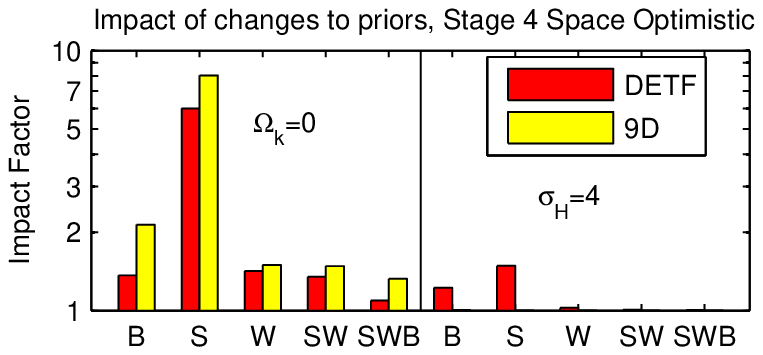}
\caption{\label{fig:DETFpriors} Impact of adding additional constraints on curvature (left side) and the
  Hubble constant (right side): 
Dark bars show the impact on the DETF FoM and light bars
  show the impact on the 9D FoM. The y-axis is the factor by which the FoM's change (for Stage 4
  space, without stage 2) when the additional constraints are
  imposed. Changing from the DETF to the 9D parameters does not
  undermine the insensitivity to additional constraints reported by
  the DETF. This plot corresponds to
  the plot on p.14 of \cite{Albrecht:2006um} (in a different format).
}
\end{figure}

Imposing additional constraints or
``priors'' on specific parameters generally will improve the figures of merit.
The DETF report has a plot similar to
Fig. \ref{fig:DETFpriors} (dark bars) which illustrates the ``impact
factor'' (factor by which the FoM improves) from imposing 
stronger priors on the curvature and the Hubble constant.
The DETF found the impact to be modest, and noted that the best data models
actually determine the parameters so well themselves
that the impact of additional priors was small.

Our work shows that DETF data models can constrain more $w(a)$
parameters than the two considered by the DETF.  
Does this improvement comes at the expense of poorer constraints on other
parameters, due to the greater flexibility of the dark-energy model?
The lighter bars in Fig. \ref{fig:DETFpriors} show the
impact factor on the 9D FoM.  In no case is the impact substantially
greater for the 9D case, indicating that going to the 9D
$w(a)$ model does not significantly undermine the constraints on the
curvature and Hubble constant from the modeled data.
In fact, the 9D impact factors are much smaller with the Hubble prior than
for the DETF FoM, suggesting that the directions in the 9D space
constrained by these data models have even less degeneracy with the Hubble parameter
than $w_0$ and $w_a$.

\section{Conclusions}

We have analyzed a FoM for dark energy probes
which is defined in a nine-dimensional parameter space, up from the
(already ambitious) two-parameter space used by the DETF.  
Our 9D FoM gives a more complete account of the impact of a given data model.
We find the DETF data models constrain significantly more
parameters than the two used by the DETF, leading to vastly
improved FoM's in the D9 space. 
Our rescaling law gives an intuitive account of the impact
of measuring more parameters.  

The 9D FoM follows the same logic as
the DETF FoM, namely it attaches equal weight to any constraint on
$w(a)$.  The measurement of a non-zero value for any of the
independently measured combinations $\vec{f}_i$ would  be equally
significant in that it would rule out a pure cosmological constant.
Thus the higher values of the 9D FoM (vs the DETF) 
reflect genuinely greater discovery power. 

The effective number of parameters measured ($D_e$) increases with the quality
of the project, and consequently so does the gap in the impact
registered by the 9D vs DETF FoM's.  Good Stage 4 projects measure
more than four parameters and produce a three or four order of
magnitude improvement in the 9D FoM vs a one order of magnitude
improvement in the DETF FoM.  
As a result, our 9D work indicates that good Stage 4 projects achieve
a ten times lower estimated cost per FoM increase than Stage 3 projects,
the reverse of the DETF conclusion.  Aside from this major difference,
the detailed DETF conclusions about the relative value of
different techniques and the importance of combining techniques are
unchanged by the 9D analysis.

\begin{acknowledgments}
We thank the DETF for a great collaboration on which this work
was based, and Lloyd Knox for his
simulated PLANCK Fisher matrices and helpful
discussions. This work was supported in part by DOE grants
DE-FG03-91ER40674 (AA) and DE-FG02-95ER40893 (GB) and by NASA
grant BEFS-04-0014-0018 (GB).   

\end{acknowledgments}

\bibliography{HDFoM}

\end{document}